\documentclass[english,conference]{IEEEtran}
\usepackage[T1]{fontenc}
\usepackage[latin9]{inputenc}
\PassOptionsToPackage{ruled, linesnumbered}{algorithm2e}
\usepackage{mathrsfs}
\usepackage{algorithm2e}
\usepackage{amsmath}
\usepackage{amsthm}
\usepackage{amssymb}
\usepackage{graphicx}

\makeatletter
\theoremstyle{plain}
\newtheorem{thm}{\protect\theoremname}
\theoremstyle{plain}
\newtheorem{lem}[thm]{\protect\lemmaname}

\usepackage{cite, balance,epstopdf}

\usepackage{amsthm}

\theoremstyle{remark}

\DeclareMathOperator{\tr}{tr}

\DeclareMathOperator{\diag}{diag}

\DeclareMathOperator{\st}{subject~to}
\newcommand{\herm}{^{{\dagger}}}
\newcommand{\trans}{^{\mathsf{T}}}

\DeclareMathOperator{\argmax}{argmax}
\makeatother

\usepackage{babel}
\providecommand{\lemmaname}{Lemma}
\providecommand{\theoremname}{Theorem}

\begin{document}
\title{\huge{Secrecy Rate Maximization for Intelligent Reflecting Surface
Assisted MIMOME Wiretap Channels}}
\author{\IEEEauthorblockN{Anshu~Mukherjee, Vaibhav~Kumar, and~Le-Nam~Tran}\IEEEauthorblockA{School of Electrical and Electronic Engineering, University College
Dublin, Belfield, Dublin 4, Ireland\\
 Email: anshu.mukherjee@ucdconnect.ie; vaibhav.kumar@ieee.org; nam.tran@ucd.ie}}
\maketitle
\begin{abstract}
Intelligent reflecting surface (IRS) has gained tremendous attention
recently as a disruptive technology for beyond 5G networks. In this
paper, we consider the problem of secrecy rate maximization for an
IRS-assisted Gaussian multiple-input multiple-output multi-antenna-eavesdropper
(MIMOME) wiretap channel (WTC). In this context, we aim to jointly
optimize the input covariance matrix and the IRS phase shifts to maximize
the achievable secrecy rate of the considered system. To solve the
formulated problem which is non-convex, we propose an iterative method
based on the block successive maximization (BSM), where \emph{each
iteration is done in closed form}. More specifically, we maximize
\emph{a lower bound} on the achievable secrecy rate to update the
input covariance matrix for fixed phase shifts, and then maximize
the (exact) achievable secrecy rate to update phase shifts for a given
input covariance. We present a convergence proof and the associated
complexity analysis of the proposed algorithm. Numerical results are
provided to demonstrate the superiority of the proposed method compared
to a known solution, and also to show the effect of different parameters
of interest on the achievable secrecy rate of the IRS-assisted MIMOME
WTC.
\end{abstract}

\section{Introduction}

\allowdisplaybreaks \sloppy

Due to the broadcast nature of radio links, wireless communications
over these channels are highly vulnerable to eavesdropping. This issue
is extremely important for military applications. Many measures have
been taken to mitigate such vulnerability. Among them, physical layer
security (PLS) has received increasing attention as one of the promising
techniques to deliver secure communication with low-complexity and
possibly (cryptographic) keyless transmission. The most fundamental
information-theoretic model for the study of PLS is so-called the
wiretap channel (WTC), where an eavesdropper aims to decode the information
intended to be exchanged between a transmitter and a legitimate receiver~\cite{MIMOME_WTC}.

On the other hand, with the recent developments in the software-controlled
hypersurface technology, it is now been possible to steer the radio
waves falling on these hypersurfaces in a controlled fashion~\cite{Ian_Magazine}.
Therefore, in order to exploit the benefits of these hypersurfaces,
termed as the intelligent reflecting surface (IRS), in the context
of PLS, the problem of secrecy rate maximization (SRM) was recently
considered in several works including~\cite{Limeng2020MIMOIRSletter,Limeng2020WTCIRS,ContDiscreteAccess2020,Zheng2021MIMOIRS}.
The IRS-assisted multiple-input multiple-output multi-antenna-eavesdropper
(MIMOME) WTC has been studied very recently in~\cite{Limeng2020MIMOIRSletter},
where in order to maximize the secrecy rate of the system under consideration,
the authors presented an alternating optimization (AO) algorithm in
combination with minorization-maximization (MM) algorithm. Also, in~\cite{Limeng2020WTCIRS}
the SRM problem for an IRS-assisted MIMOME WTC was studied for both
with and without the knowledge of eavesdropper's channel, where the
authors proposed an AO-based algorithm. In~\cite{ContDiscreteAccess2020},
the SRM problem for IRS-assisted MIMOME WTC with both continuous as
well as discrete phase shifts at the IRS was considered, where a successive
convex approximation (SCA) based AO algorithm was used to find a suboptimal
solution.

In this paper, we propose an efficient algorithm to find the input
convariance matrix and the IRS phase shifts to maximize the achievable
secrecy rate of the IRS-assisted MIMOME WTC using the block successive
maximization (BSM) framework \cite{razaviyayn2012BSM}. More specifically,
a lower bound on the achievable secrecy rate is considered when optimizing
the input convariance matrix and exact maximization is performed for
each individual phase shift. In particular, these optimization steps
are done using closed-form expressions. We compare the convergence
speed and average run time of the proposed solution with an existing
method to establish the superiority of our proposed method. Extensive
numerical experiments are also carried out to demonstrate the effect
of different system parameters, such as the number of reflecting elements
at the IRS, the number of receive antennas at the Eavesdropper and
the power transmitted from the transmitter, on the achievable secrecy
rate of the considered IRS-assisted MIMOME WTC.

\textit{Notation:} In this paper, we use bold uppercase and lowercase
letters to denote matrices and vectors, respectively. $(\cdot)\herm$,
$(\cdot)\trans$ and $(\cdot)^{*}$ represent the Hermitian transpose,
ordinary transpose and conjugate operators, respectively. We use $\mathbb{C}^{M\times N}$
to denote the space of $M\times N$ complex matrices. By $\mathbf{Y}_{i,j}$
we represent the $j$-th element of $i$-th row of matrix $\mathbf{Y}$;
$\diag(\mathbf{y})$ denotes the diagonal square matrix whose (main)
diagonal elements are taken from $\mathbf{y}$. $\mathbf{I}$ and
$\mathbf{0}$ specify identity and zero matrices respectively, the
size of which can be easily inferred from the context. We denote the
trace and determinant of the matrix $\mathbf{Y}$ by $\tr(\boldsymbol{\mathbf{Y}})$
and $\left|\mathbf{Y}\right|$, respectively. Furthermore, we represent
the expected value of a random variable by $\mathbb{E}[\cdot]$ and
the real part of a complex number by $\Re\left\{ \cdot\right\} $.
For $\mathbf{y}\in\mathbb{R}^{N}$, $[\mathbf{y}]_{+}=\begin{bmatrix}\max(y_{1},0), & \max(y_{2},0), & \cdots & \max(y_{N},0)\end{bmatrix}$,
where $\mathbb{R}$ denotes the set of real numbers. By $\mathbf{A}\succeq(\textrm{resp.}\ \succ)\ \mathbf{B}$
we mean $\mathbf{A}-\mathbf{B}$ is positive semidefinite (resp. definite).
$|x|$ and $\angle x$ denote the modulus and the phase of a complex
number $x$.

\section{System Model and Problem Formulation}

In this section, we describe the system model and formulate the problem
of maximizing the achievable secrecy rate for the system under consideration
.

\subsection{System Model}

\begin{figure}
\begin{centering}
\includegraphics[width=0.75\columnwidth]{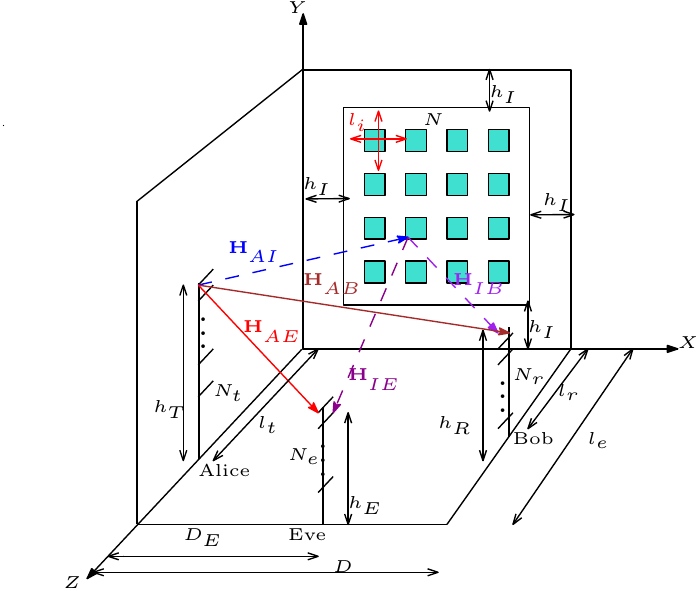}
\par\end{centering}
\caption{A block diagram of IRS-assisted MIMOME WTC system.}
\label{fig:SystemModel}
\end{figure}
Let us consider an IRS-aided MIMOME WTC system where Alice is the
transmitter, Bob is the (legitimate) receiver, and Eve is the eavesdropper.
The numbers of antennas at Alice, Bob and Eve are $N_{t}$, $N_{r}$,
and $N_{e}$, respectively, and the IRS is made-up of $N$ low-cost
passive reflecting elements. The system model is shown in~Fig.~\ref{fig:SystemModel}.
The locations of Alice, Bob and Eve in 3-dimensional Euclidean space
are $(0,0,l_{t})$, $(D,0,l_{r})$ and $(D_{E},0,l_{e})$, respectively.
The height (measured from the $X$-Z plane) of the top-most antenna
at Alice, Bob and Eve is respectively given by $h_{T}$, $h_{R}$,
and $h_{E}$. The distance between each antenna at Alice is denoted
by $\mathscr{\iota}_{a}$, and that between each antenna at Bob and
Eve are respectively denoted by $\mathscr{\iota}_{b}$ and $\mathscr{\iota}_{e}$.
The distance of an IRS element to its immediate neighboring one is
denoted by $\iota_{i}$. The complex-valued channel matrices for the
Alice-IRS, IRS-Bob, IRS-Eve, Alice-Bob and Alice-Eve links are denoted
by $\mathbf{H}_{AI}\in\mathbb{C}^{N\times N_{t}}$, $\mathbf{H}_{IB}\in\mathbb{C}^{N_{r}\times N}$,
$\mathbf{H}_{IE}\in\mathbb{C}^{N_{e}\times N}$, $\mathbf{H}_{AB}\in\mathbb{C}^{N_{b}\times N_{t}}$,
and $\mathbf{H}_{AE}\in\mathbb{C}^{N_{e}\times N_{t}}$, respectively.
It is assumed that all these channel matrices are quasi-static and
perfectly known at all of the nodes.

The received signals at Bob and Eve are, respectively, expressed as
\begin{equation}
\begin{aligned}\mathbf{y}_{b}= & \mathbf{H}_{AB}\mathbf{x}+\mathbf{H}_{IB}\mathbf{\mathbf{Z}(\boldsymbol{\theta})}\mathbf{H}_{AI}\mathbf{x}+\mathbf{n}_{b},\\
\mathbf{y}_{e}= & \mathbf{H}_{AE}\mathbf{x}+\mathbf{H}_{IE}\mathbf{Z}(\boldsymbol{\theta})\mathbf{H}_{AI}\mathbf{x}+\mathbf{n}_{e},
\end{aligned}
\label{eq:RxSignals}
\end{equation}
where $\mathbf{x}\in\mathbb{C}^{N_{t}\times1}$ represents the transmitted
signal from Alice; $\mathbf{n}_{b}\sim\mathcal{CN}(\mathbf{0},\sigma_{b}^{2}\mathbf{I})$
and $\mathbf{n}_{e}\sim\mathcal{CN}(\mathbf{0},\sigma_{e}^{2}\mathbf{I})$
are the additive white Gaussian noise at Bob and Eve, respectively.
In (\ref{eq:RxSignals}), $\mathbf{Z}(\boldsymbol{\theta})\triangleq\diag(\boldsymbol{\theta})$,
where $\boldsymbol{\theta}\triangleq[\theta_{1},\theta_{2},\dots,\theta_{N}]\trans\in\mathbb{C}^{N\times1}$,
$\theta_{i}=e^{j\phi_{i}}$, $i\in\mathcal{I}\triangleq\left\{ 1,2,\ldots,N\right\} $,
and $\phi_{i}\in[0,2\pi)$ denotes the phase shift induced by the
$i$-th reflecting element at the IRS.

\subsection{Problem Formulation}

Let $\mathbf{X}\triangleq\mathbb{E}\left\{ \mathbf{x}\mathbf{x}\herm\right\} \succeq\mathbf{0}$
be the input covariance matrix. Then for given $\mathbf{X}$ and $\boldsymbol{\theta}$,
the following secrecy rate (in nat/s/Hz) between Alice and Bob is
achievable~(c.f.~\cite{Oggier2011SecCapEq})
\begin{equation}
C_{s}(\boldsymbol{\theta},\mathbf{X})=[C_{B}(\boldsymbol{\theta},\mathbf{X})-C_{E}(\boldsymbol{\theta},\mathbf{X})]_{+},\label{eq:SecrecyRateDef}
\end{equation}
where $C_{B}(\mathbf{X})$ and $C_{E}(\mathbf{X})$ denote the achievable
rate at Bob and Eve, respectively, defined as 
\begin{subequations}
\begin{align}
C_{B}(\boldsymbol{\theta},\mathbf{X}) & =\ln\bigl|\mathbf{I}+\mathbf{H}_{B}(\boldsymbol{\theta})\mathbf{X}\mathbf{H}_{B}\herm(\boldsymbol{\theta})\bigr|\label{eq:BobSignal}\\
C_{E}(\boldsymbol{\theta},\mathbf{X}) & =\ln\bigl|\mathbf{I}+\mathbf{H}_{E}(\boldsymbol{\theta})\mathbf{X}\mathbf{H}_{E}\herm(\boldsymbol{\theta})\bigr|\label{eq:EveSignal}
\end{align}
\end{subequations}
where $\mathbf{H}_{B}(\boldsymbol{\theta})\triangleq\bar{\mathbf{H}}_{AB}+\bar{\mathbf{H}}_{IB}\mathbf{Z}(\boldsymbol{\theta})\mathbf{H}_{AI}$,
$\mathbf{H}_{E}(\boldsymbol{\theta})\triangleq\bar{\mathbf{H}}_{AE}+\bar{\mathbf{H}}_{IE}\mathbf{Z}(\boldsymbol{\theta})\mathbf{H}_{AI}$,
$\bar{\mathbf{H}}_{AB}\triangleq\tfrac{1}{\sigma_{b}}\mathbf{H}_{AB}$,
$\bar{\mathbf{H}}_{IB}\triangleq\tfrac{1}{\sigma_{b}}\mathbf{H}_{IB}$,
$\bar{\mathbf{H}}_{AE}\triangleq\tfrac{1}{\sigma_{e}}\mathbf{H}_{AE}$
and $\bar{\mathbf{H}}_{IE}\triangleq\tfrac{1}{\sigma_{e}}\mathbf{H}_{IE}$.

The problem of SRM under the sum power constraint (SPC) reads
\begin{align}
\underset{\mathbf{X}\in\mathcal{X}\,\boldsymbol{\theta}\in\Theta}{\mathrm{maximize}} & \ C_{s}(\boldsymbol{\theta},\mathbf{X})\label{eq:MainProblem}
\end{align}
where
\begin{equation}
\begin{aligned}\mathcal{X} & =\left\{ \mathbf{X}:\mathbf{X}\succeq\mathbf{0}\ |\ \tr(\mathbf{X})\leq P_{0}\right\} ,\\
\Theta & =\left\{ \boldsymbol{\theta}:\left|\theta_{i}\right|=1,\forall i\in\mathcal{I}\right\} ,
\end{aligned}
\label{eq:FeasibleSetsDef}
\end{equation}
and $P_{0}$ is the maximum power budget at Alice. Note that in~(\ref{eq:FeasibleSetsDef}),
$\tr(\mathbf{X})\leq P_{0}$ denotes the transmit power constraint
and $\left|\theta_{i}\right|=1$ denotes the unit-modulus constraint.

To appreciate the novelty of our proposed method presented in the
next section for solving (\ref{eq:MainProblem}), we discuss the drawbacks
of existing solutions known to us for solving the same problem. As
mentioned previously in the introduction section, an AO-based algorithm
was presented in~\cite{Limeng2020WTCIRS} that alternately optimizes
$\mathbf{X}$ and $\boldsymbol{\theta}$ while the other variable
is fixed. More specifically, a barrier method was developed to find
$\mathbf{X}$ for a given $\boldsymbol{\theta}$, and then for a given
$\mathbf{X}$, each $\theta_{i}$ was found using Dinkelbach's method.
Similarly, an AO-like algorithm was proposed in~\cite{Jiang2020MIMOIRS}
but an SCA method was derived for finding $\mathbf{X}$ for a given
$\boldsymbol{\theta}$, and then each $\theta_{i}$ is optimized sequentially
using a linear search procedure. It can easily be noted that both
the methods mentioned above incur high complexity to produce a solution.
Motivated by this, in the next section, we propose an efficient method
based on the BSM to find a stationary solution to problem~(\ref{eq:MainProblem}).

\section{Closed-form Design based on Block Successive Maximization}

\subsection{Algorithm Description}

The proposed method is based on the BSM method \cite{razaviyayn2012BSM}
where $\mathbf{X}$ and each $\theta_{i}$ are viewed as individual
blocks. We note that the main principle of the BSM is that a single
block is updated in each iteration using a proper bound or exact optimization.
The BSM method is particularly efficient if the optimization at each
step is computationally cheap. To this end we propose an iterative
method as follows:
\begin{itemize}
\item We sequentially update each phase shift $\theta_{i}$, while other
variables are fixed and \emph{exact optimization} is considered. We
derive closed-form expressions for this step, rather than using Dinkelbach's
method \cite{Limeng2020WTCIRS} or a line search \cite{Jiang2020MIMOIRS}.
\item We update $\mathbf{X}$ using a lower bound that leads to a water-filling-like
solution. This step is different from the SCA-based method in \cite{Jiang2020MIMOIRS}
where the lower bound is repeatedly solved. In contrast we only maximize
the lower bound \emph{once} in each iteration. Despite this, the proposed
method is provably convergent to a stationary point of (\ref{eq:MainProblem}).
\end{itemize}
The details of the proposed algorithm are given in the following subsections.

\subsubsection{Optimizing $\theta_{i}$ for given $\mathbf{X}$ and other phase
shifts $\theta_{j},j\in\left\{ \mathcal{I}\setminus i\right\} $}

The optimization of each $\theta_{i}$, while remaining $\theta_{j},j\in\left\{ \mathcal{I}\setminus i\right\} $
and $\mathbf{X}$ are held fixed, is formulated as \begin{subequations}
\label{eq:CsbarTheta}
\begin{align}
\underset{\theta_{i}}{\mathrm{maximize}} & \ \bar{C}_{s}(\theta_{i})\\
\st & \ \left|\theta_{i}\right|=1,
\end{align}
 \end{subequations}where
\begin{multline}
\bar{C}_{s}(\theta_{i})=\ln\Bigl|\mathbf{I}+\theta_{i}\mathbf{P}_{i}^{-1}\mathbf{Q}_{i}+\theta_{i}^{*}\mathbf{P}_{i}^{-1}\mathbf{Q}_{i}\herm\Bigr|\\
-\ln\Bigl|\mathbf{I}+\theta_{i}\mathbf{R}_{i}^{-1}\mathbf{S}_{i}+\theta_{i}^{*}\mathbf{R}_{i}^{-1}\mathbf{S}_{i}\herm\Bigr|,\label{eq:EqPQRS}
\end{multline}
and detailed expressions for $\mathbf{P}_{i}$, $\mathbf{Q}_{i}$,
$\mathbf{R}_{i}$ and $\mathbf{S}_{i}$ are given in ~Appendix~\ref{sec:APP-Closed-form-solution}.
We now derive a closed-solution for (\ref{eq:EqPQRS}) for the non-trivial
case where $\tr(\mathbf{P}_{i}^{-1}\mathbf{Q}_{i})\ne0$ and $\tr(\mathbf{R}_{i}^{-1}\mathbf{S}_{i})\ne0$,
and refer the interested readers to \cite{Limeng2020WTCIRS} for the
trivial cases where $\tr(\mathbf{P}_{i}^{-1}\mathbf{Q}_{i})=0$ and/or
$\tr(\mathbf{R}_{i}^{-1}\mathbf{S}_{i})=0$.

Let $\mathbf{P}_{i}^{-1}\mathbf{Q}_{i}=\tilde{\mathbf{U}}_{ib}\tilde{\boldsymbol{\Sigma}}_{ib}\tilde{\mathbf{U}}_{ib}\herm$
be the eigenvalue decomposition (EVD) of $\mathbf{P}_{i}^{-1}\mathbf{Q}_{i}$.
Since $\mathbf{Q}_{i}$ is a rank-1 matrix we can write $\tilde{\boldsymbol{\Sigma}}_{ib}=\diag([\gamma_{ib},0,\ldots,0]\trans)$
where $\gamma_{ib}$ is only non-zero eigenvalue of $\mathbf{P}_{i}^{-1}\mathbf{Q}_{i}$.
Similarly, let $\tilde{\mathbf{U}}_{ie}\tilde{\boldsymbol{\Sigma}}_{ie}\tilde{\mathbf{U}}_{ie}\herm=\mathbf{R}_{i}^{-1}\mathbf{S}_{i}$
be the EVD of $\mathbf{R}_{i}^{-1}\mathbf{S}_{i}$ where $\tilde{\boldsymbol{\Sigma}}_{ie}=\diag([\gamma_{ie},0,\ldots,0]\trans)$
and $\gamma_{ie}$ is only non-zero eigenvalue of $\mathbf{R}_{i}^{-1}\mathbf{S}_{i}$.
Furthermore, let $\mathbf{A}_{i}=\tilde{\mathbf{U}}_{ib}\herm\mathbf{P}_{i}\tilde{\mathbf{U}}_{ib}$,
and $\bar{\mathbf{a}}_{i}$ denotes the first column of $\mathbf{A}_{i}^{-1}$,
and $\tilde{\mathbf{a}}_{i}\trans$ denotes the first row of $\mathbf{A}_{i}$.
Similarly, we define $\mathbf{B}_{i}=\tilde{\mathbf{U}}_{ie_{i}}\herm\mathbf{R}_{i}\tilde{\mathbf{U}}_{ie}$,
and $\bar{\mathbf{b}}_{i}$ denotes the first column of $\mathbf{B}_{i}^{-1}$,
and $\tilde{\mathbf{b}}_{i}\trans$ denotes the first row of $\mathbf{B}_{i}$.
We can further rewrite~(\ref{eq:EqPQRS}) as~(c.f.~\cite{Zhang2020MIMOIRS})
\begin{equation}
\bar{C}_{s}(\theta_{i})=\ln\left(\frac{2\Re(\gamma_{ib}\theta_{i})+\delta_{ib}}{2\Re(\gamma_{ie}\theta_{i})+\delta_{ie}}\right),\label{eq:C_bar_reform}
\end{equation}
where $\delta_{ib}=1+|\gamma_{ib}|^{2}(1-\bar{a}_{i1}\tilde{a}_{i1})$,
$\delta_{ie}=1+|\gamma_{ie}|^{2}(1-\bar{b}_{i1}\tilde{b}_{i1})$,
and $\bar{a}_{i1}$ and $\tilde{a}_{i1}$ denote the first element
of $\bar{\mathbf{a}}_{i}$ and $\tilde{\mathbf{a}}_{i}\trans$, respectively.
It is easy to see that using~(\ref{eq:C_bar_reform}), problem~(\ref{eq:CsbarTheta})
is equivalent to \begin{subequations}\label{eq:thetaSolveobj}
\begin{align}
\underset{\theta_{i}}{\mathrm{maximize}} & \ \frac{2\Re(\gamma_{ib}\theta_{i})+\delta_{ib}}{2\Re(\gamma_{ie}\theta_{i})+\delta_{ie}}\\
\st & \ \left|\theta_{i}\right|=1.
\end{align}
 \end{subequations}Let $\phi_{ib}=\angle\gamma_{ib}$, $\phi_{ie}=\angle\gamma_{ie}$
and $\lambda_{ib}=2$$\left|\gamma_{ib}\right|$ and $\lambda_{ie}=2$$\left|\gamma_{ie}\right|$.
Then~(\ref{eq:thetaSolveobj}) is equivalent to \begin{subequations}\label{eq:thetaTrigonoObj}
\begin{align}
\underset{\phi_{i}}{\mathrm{maximize}} & \ \frac{\lambda_{ib}\cos(\phi_{i}-\phi_{ib})+\delta_{ib}}{\lambda_{ie}\cos(\phi_{i}-\phi_{ie})+\delta_{ie}}=f(\phi_{i})\\
\st & \ 0\leq\phi_{i}<2\pi.
\end{align}
\end{subequations} The derivative of the objective function in (\ref{eq:thetaTrigonoObj})
is given by 
\begin{align}
f^{'}(\phi_{i}) & =\frac{\lambda_{ib}\lambda_{ie}\sin(\phi_{ib}-\phi_{ie})-\lambda_{i}\sin(\phi_{i}-\psi_{i})}{\left(\lambda_{ie}\cos(\phi_{i}-\phi_{ie})+\delta_{ie}\right)^{2}},
\end{align}
where 
\begin{align}
\lambda_{i} & \triangleq\sqrt{\lambda_{ib}^{2}\delta_{ie}^{2}+\lambda_{ie}^{2}\delta_{ib}^{2}-2\lambda_{ib}\lambda_{ie}\delta_{ib}\delta_{ie}\cos(\phi_{ib}-\phi_{ie})},\\
\psi_{i} & \triangleq\arctan\left\{ \frac{-\lambda_{ib}\delta_{ie}\sin(\phi_{ib})+\lambda_{ie}\delta_{ib}\sin(\phi_{ie})}{\lambda_{ib}\delta_{ie}\cos(\phi_{ib})-\lambda_{ie}\delta_{ib}\cos(\phi_{ie})}\right\} .
\end{align}
We note that $\delta_{iu}\geq1+\lambda_{iu}$ for $u\in\{b,e\}$,
which holds due to their definitions. Thus, the equation $f^{'}(\phi_{i})=0$
has two possible solutions as follows:
\begin{align}
\phi_{i_{1}} & =\arcsin\left\{ \frac{\lambda_{ib}\lambda_{ie}}{\lambda_{i}}\sin(\phi_{ib}-\phi_{ie})\right\} +\psi_{i},\label{eq:Phi_1}\\
\phi_{i_{2}} & =\pi-\arcsin\left\{ \frac{\lambda_{ib}\lambda_{ie}}{\lambda_{i}}\sin(\phi_{ib}-\phi_{ie})\right\} +\psi_{i}.\label{eq:Phi_2}
\end{align}
Thus, the optimal solution to~(\ref{eq:CsbarTheta}) is found as

\begin{equation}
\theta_{i}^{\star}=e^{j\phi_{i}^{\star}}\label{eq:ThetaClsdFrmSol}
\end{equation}
where\footnote{We can also check the second derivative of these three critical points
to find the optimal solution but comparing their objective values
is much simpler.}
\begin{equation}
\phi_{i}^{\star}=\arg\max\{f(0),f(\phi_{i_{1}}),f(\phi_{i_{2}})\}.\label{eq:Phi_Opt}
\end{equation}
\begin{algorithm}
\caption{Block Successive Maximization Method}

\label{alg:BSM}

\KwIn{ $\mathbf{\mathbf{X}}_{0}\ensuremath{\in}\mathcal{X}$, $\boldsymbol{\theta}_{0}\in\Theta$}

\KwOut{$\mathbf{\mathbf{X}}_{k}\ensuremath{\in}\mathcal{X}$, $\boldsymbol{\theta}_{k}\in\Theta$}

$k\leftarrow1$\;

\Repeat{convergence }{

\For{$i\in\mathcal{I}$}{\label{alg:BSM:thetaupdateSTART}

Compute $\theta_{i}$ according to (\ref{eq:ThetaClsdFrmSol}) for
fixed $\mathbf{X}_{k-1}$\label{alg:BSM:thetaupdate}

$\theta_{i}\leftarrow\theta_{i}^{\star}$

}\label{alg:BSM:thetaupdateEND}

Find $\mathbf{X}_{k}$ for fixed $\boldsymbol{\theta}_{k}$ according
to (\ref{eq:FindXsca})\label{alg:BSM:Xupdate}

$k\leftarrow k+1$

}
\end{algorithm}

\subsubsection{Optimizing $\mathbf{X}$ for given $\boldsymbol{\theta}$}

The next step is to optimize $\mathbf{X}$ for a given $\boldsymbol{\theta}$.
Instead of maximizing the secrecy rate exactly, we consider a lower
bound in this step. Let $\mathbf{X}_{k-1}$ be the value of $\mathbf{X}$
at iteration $k-1$. Then, due to the concavity of the term $\ln|\cdot|$
it follows that 
\begin{align}
 & C(\boldsymbol{\theta}_{k},\mathbf{X})\leq\hat{C}_{s}(\boldsymbol{\theta}_{k},\mathbf{X})=\ln\bigl|\mathbf{I}+\mathbf{H}_{B}(\boldsymbol{\theta}_{k})\mathbf{X}\mathbf{H}_{B}\herm(\boldsymbol{\theta}_{k})\bigr|\nonumber \\
 & -\ln\bigl|\mathbf{I}+\mathbf{H}_{E}(\boldsymbol{\theta}_{k})\mathbf{X}_{k-1}\mathbf{H}_{E}\herm(\boldsymbol{\theta}_{k})\bigr|-\tr\left(\bigl(\boldsymbol{\Phi}_{k-1}\bigr)(\mathbf{X}-\mathbf{X}_{k-1})\right),\label{eq:ConvX1}
\end{align}
where $\boldsymbol{\Phi}_{k-1}=\mathbf{H}_{E}\herm(\boldsymbol{\theta})\bigl(\mathbf{I}+\mathbf{H}_{E}(\boldsymbol{\theta})\mathbf{X}_{k-1}\mathbf{H}_{E}\herm(\boldsymbol{\theta})\bigr)^{-1}\mathbf{H}_{E}(\boldsymbol{\theta}).$
Note that (\ref{eq:ConvX1}) is obtained by using a first-order approximation
of the term $\ln\left|\mathbf{I}+\mathbf{H}_{E}(\boldsymbol{\theta})\mathbf{X}\mathbf{H}_{E}\herm(\boldsymbol{\theta})\right|$
around $\mathbf{X}_{k-1}$. Next, we update $\mathbf{X}_{k}$ as $\mathbf{X}_{k}=\argmax\{\hat{C}_{s}(\boldsymbol{\theta}_{k},\mathbf{X})\ |\ \mathbf{X}\in\mathcal{X}\}$
which is equivalent to \begin{subequations}\label{eq:FindXsca}
\begin{align}
\underset{\mathbf{X}\succeq\mathbf{0}}{\mathrm{maximize}} & \ \ln\bigl|\mathbf{I}+\mathbf{H}_{B}(\boldsymbol{\theta})\mathbf{X}\mathbf{H}_{B}\herm(\boldsymbol{\theta})\bigr|-\tr\bigl(\boldsymbol{\Phi}_{k-1}\mathbf{X}\bigr)\label{eq:LinearlizedX}\\
\st & \ \tr(\mathbf{X})\leq P_{0}.\label{eq:SPC}
\end{align}
\end{subequations}The aforementioned problem~(\ref{eq:FindXsca})
admits a \emph{water-filling solution~}\cite{ThangNguyen2020}. To
lighten the notations we write $\mathbf{H}_{B}$ instead of $\mathbf{H}_{B}(\boldsymbol{\theta})$.
Let $\mu\geq0$ be the Lagrangian multiplier of (\ref{eq:SPC}). Then
the \emph{partial} Lagrangian multiplier of (\ref{eq:FindXsca}) is
\begin{align}
\mathcal{L}(\mathbf{X},\mu) & =\ln|\mathbf{I}+\mathbf{H}_{B}\mathbf{X}\mathbf{H}_{B}\herm|-\tr\bigl(\boldsymbol{\Phi}_{k-1}\mathbf{X}\bigr)-\mu\bigl(\tr(\mathbf{X})-P_{0}\bigr)\nonumber \\
 & =\ln|\mathbf{I}+\mathbf{H}_{B}\mathbf{X}\mathbf{H}_{B}\herm|-\tr\bigl(\bar{\boldsymbol{\Phi}}_{k-1}\mathbf{X}\bigr)+\mu P_{0}.
\end{align}
where $\bar{\boldsymbol{\Phi}}_{k-1}=\boldsymbol{\Phi}_{k-1}+\mu\mathbf{I}$.
The dual function is given by 
\begin{equation}
g(\mu)=\underset{\mathbf{X}\succeq\mathbf{0}}{\max}\ \ln|\mathbf{I}+\mathbf{H}_{B}\mathbf{X}\mathbf{H}_{B}\herm|-\tr\bigl(\bar{\boldsymbol{\Phi}}_{k-1}\mathbf{X}\bigr)+\mu P_{0}.\label{eq:WF-1}
\end{equation}
To evaluate the dual function for a given $\mu$, let $\bar{\mathbf{X}}\triangleq\bar{\boldsymbol{\Phi}}_{k-1}^{1/2}\mathbf{X}\bar{\boldsymbol{\Phi}}_{k-1}^{1/2}$.
Then the above maximization is equivalent to
\begin{equation}
\underset{\bar{\mathbf{X}}\succeq\mathbf{0}}{\max}\ \mathcal{L}(\bar{\mathbf{X}},\mu)=\ln\bigl|\mathbf{I}+\mathbf{H}_{B}\bar{\boldsymbol{\Phi}}_{k-1}^{-1/2}\bar{\mathbf{X}}\bar{\boldsymbol{\Phi}}_{n-1}^{-1/2}\mathbf{H}_{B}\herm\bigr|-\tr(\bar{\mathbf{X}}).\label{eq:WF-3}
\end{equation}
Denote the eigenvalue decomposition (EVD) of $\bar{\boldsymbol{\Phi}}_{k-1}^{-1/2}\mathbf{H}_{B}\herm\mathbf{H}_{B}\bar{\boldsymbol{\Phi}}_{n-1}^{-1/2}$
by $\mathbf{U}\boldsymbol{\Sigma}\mathbf{U}\herm$ where $\mathbf{U}\in\mathbb{C}^{N_{t}\times N_{t}}$
is unitary, $\boldsymbol{\Sigma}=\diag(\sigma_{1},\sigma_{2},\dots,\sigma_{r},0,\ldots,0)$,
and $r$ is the rank of $\bar{\boldsymbol{\Phi}}_{n-1}^{-1/2}\mathbf{H}_{B}$
and let $\dot{\mathbf{X}}=\mathbf{U}\herm\bar{\mathbf{X}}\mathbf{U}$.
Then (\ref{eq:WF-3}) is further equivalent to
\begin{equation}
\underset{\dot{\mathbf{X}}\succeq\mathbf{0}}{\max}\ \mathcal{L}(\dot{\mathbf{X}},\mu)=\ln\bigl|\mathbf{I}+\boldsymbol{\Sigma}\dot{\mathbf{X}}\bigr|-\tr(\dot{\mathbf{X}}).\label{eq:WF-4}
\end{equation}
It is now easy to see that we can assume $\dot{\mathbf{X}}$ to be
diagonal (due to Hadamard's inequality), and the optimal solution
to~(\ref{eq:WF-4}) is given by $\mathbf{\dot{\mathbf{X}}=\diag}\bigl([1-\frac{1}{\sigma_{1}}]_{+},\ldots,[1-\frac{1}{\sigma_{r}}]_{+},0,\ldots,0\bigr)$.
In summary, the optimal solution to (\ref{eq:WF-1}) is
\begin{equation}
\mathbf{X}=\bar{\boldsymbol{\Phi}}_{k-1}^{-1/2}\mathbf{U}\dot{\mathbf{X}}\mathbf{U}\herm\bar{\boldsymbol{\Phi}}_{k-1}^{-1/2}.
\end{equation}
The next step is to solve the dual problem $\min\ \{g(\mu)\ |\ \mu\geq0$
which can be done efficiently using a bisection search. We refer the
interested readers to\cite{ThangNguyen2020} for further details.

The overall algorithm to solve~(\ref{eq:MainProblem}) is summarized
in~Algorithm~\ref{alg:BSM}.

\subsection{Convergence Analysis}

We now show that Algorithm~\ref{alg:BSM} indeed converges to a stationary
point of (\ref{eq:MainProblem}). In particular the following lemma
is in order
\begin{lem}
\label{lemma:converge}The objective sequence generated by Algorithm~\ref{alg:BSM}
is monotonically increasing, i.e., 
\begin{equation}
C_{s}(\boldsymbol{\theta}_{k},\mathbf{\mathbf{X}}_{k})\geq C_{s}(\boldsymbol{\theta}_{k},\mathbf{\mathbf{X}}_{k-1})\geq C_{s}(\boldsymbol{\theta}_{k-1},\mathbf{\mathbf{X}}_{k-1}).\label{eq:CSequence}
\end{equation}
Also, the iterate sequence $(\boldsymbol{\theta}_{k},\mathbf{\mathbf{X}}_{k})$
converges to a limit point which is a stationary solution of (\ref{eq:MainProblem}).
\end{lem}
\begin{IEEEproof}
See Appendix~\ref{sec:Proof-of-Proposition}.
\end{IEEEproof}

\subsection{Complexity Analysis}

In this subsection we provide the complexity analysis of Algorithm~\ref{alg:BSM}.
In particular we adopt the big-$\mathcal{O}$ notation and present
the number of complex multiplications for each iteration of Algorithm~\ref{alg:BSM}.
To compute $\mathbf{P}_{i}$ we use~(\ref{eq:Pi:simple}) and note
that $\mathbf{P}$ only needs to be computed once for all $\mathbf{P}_{i}$.
The complexity to obtain $\mathbf{P}$ is $\mathcal{O}(NN_{r}^{2})$
and to compute each $\mathbf{P}_{i}$ we require $\mathcal{O}(N_{t}^{2}N_{r})$
additional complex multiplications. In the same way, the complexity
to compute each $\mathbf{R}_{i}$ is $\mathcal{O}(N_{t}^{2}N_{e})$.
We skip the complexity of obtaining $\mathbf{Q}_{i}$ and $\mathbf{S}_{i}$
since it is much less than that of obtaining $\mathbf{P}_{i}$ and
$\mathbf{R}_{i}$. It is easy to see that the complexity of computing
$\mathbf{P}_{i}^{-1}\mathbf{Q}_{i}$ and its EVD is $\mathcal{O}(N_{r}^{3})$.
Similarly, the complexity of computing $\mathbf{R}_{i}^{-1}\mathbf{S}_{i}$
and its EVD is $\mathcal{O}(N_{e}^{3})$. The complexity of computing
$\bar{a}_{i1}$, $\tilde{a}_{i1}$, $\bar{b}_{i1}$, and $\tilde{b}_{i1}$
, the closed-form expressions for each optimal $\theta_{i}^{\star}$
is much less than that of computing other terms, and thus is omitted.
When $\boldsymbol{\theta}$ is fixed, it can be shown that the complexity
for solving~(\ref{eq:LinearlizedX}) is $\mathcal{O}(N_{e}^{3}+N_{t}^{2}N_{e}+N_{t}N_{e}^{2}+N_{t}^{3})$.
In summary, the per-iteration complexity of Algorithm \ref{alg:BSM}
is $\mathcal{O}(N(N_{r}^{2}+N_{t}^{2}N_{r}+N_{t}^{2}N_{e}+N_{r}^{3}+N_{e}^{3})+N_{t}N_{e}^{2}+N_{t}^{3})$.

\begin{figure*} 
\centering 
	\begin{minipage}{.32\textwidth}   
		\centering   
		\includegraphics[width=.99\linewidth,height=4.5cm]{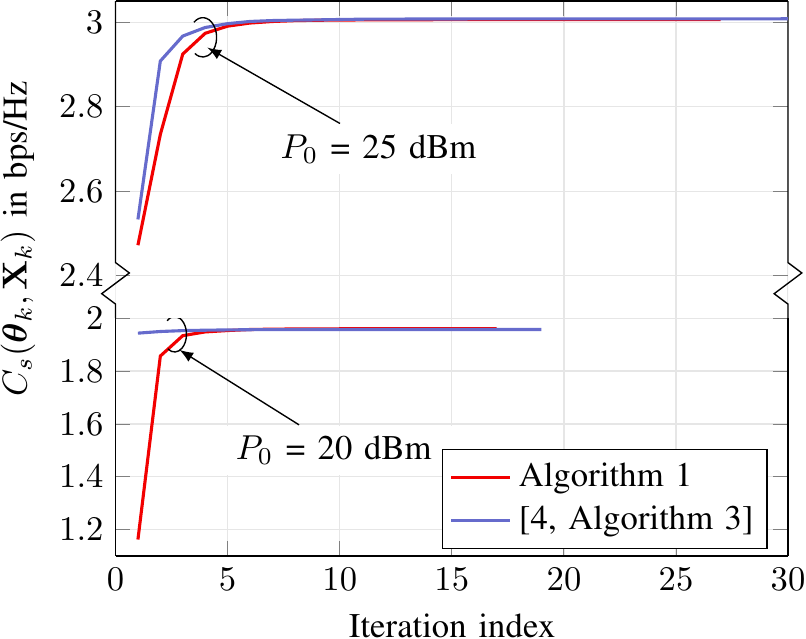}   
		\caption{Convergence comparison for $(N_t,N_r,N_e, N) = (4,3,2,25).$}   
		\label{fig2Convergence} 
	\end{minipage}%
\hfill
	\begin{minipage}{.32\textwidth}   
		\centering   
		\includegraphics[width=.99\linewidth,height=4.5cm]{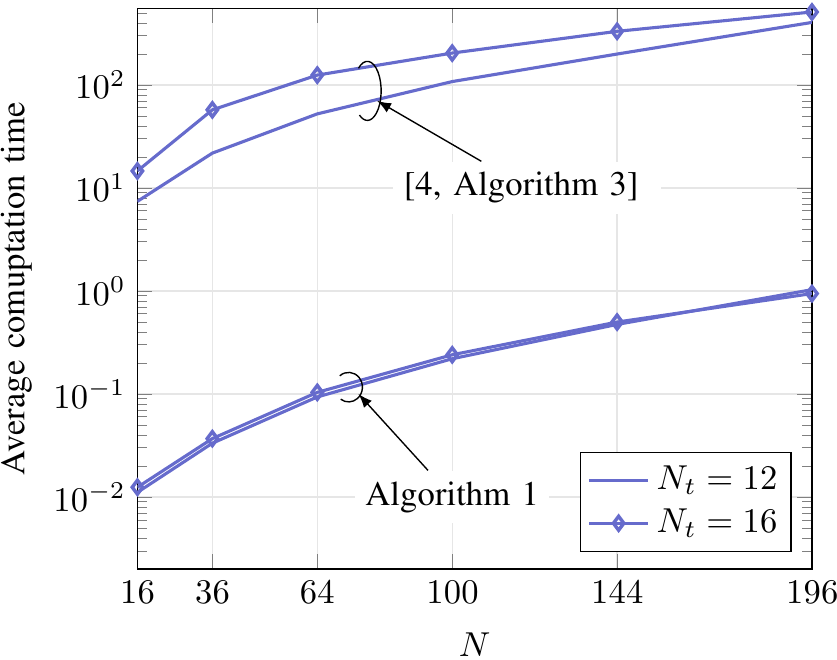}   
		\caption{Average computation time for various algorithms for $(N_r, N_e) = (8,6)$ and $P_0 = 10$~dB.}   
		\label{figThetaTimeCheck} 
	\end{minipage} 
\hfill
	\begin{minipage}{.32\textwidth}   
		\centering   
		\includegraphics[width=.99\linewidth,height=4.5cm]{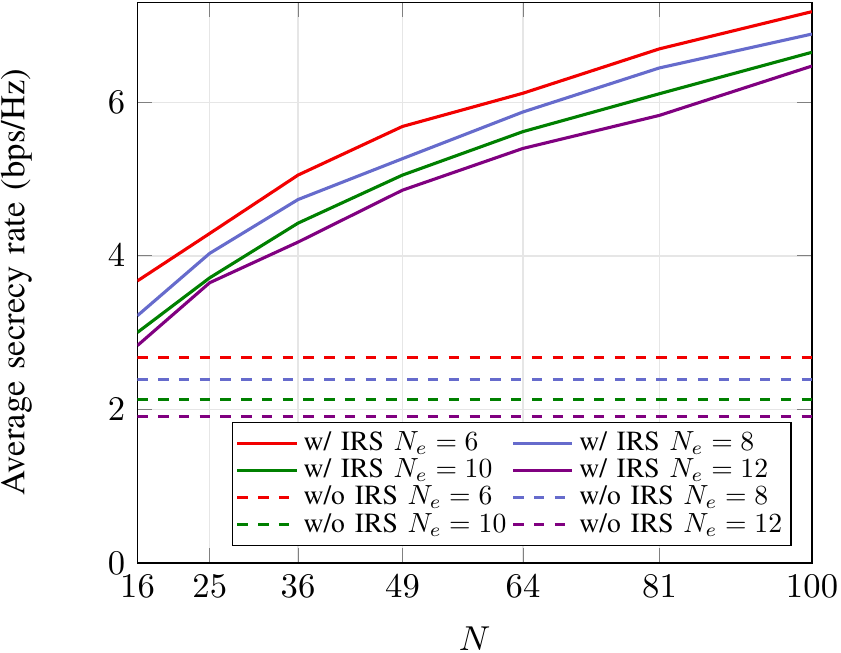}   
		\caption{The average secrecy rate with $(N_{t},N_{r})=\text{(12,8)}$ and $P_{0}=25$ dBm.}   
		\label{fig3IRSimpact} 
	\end{minipage} 
\end{figure*}

\begin{figure}
\begin{centering}
\includegraphics[viewport=62bp 559bp 285bp 741bp,width=0.7\columnwidth]{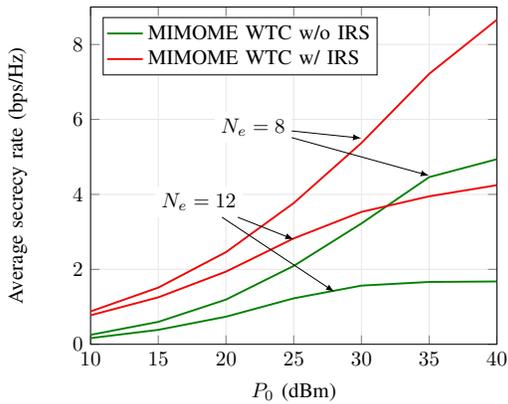}
\par\end{centering}
\caption{Comparison of the secrecy rate for MIMOME WTC and IRS-assisted MIMOME
WTC for $(N_{t},N_{r},N)=(8,6,25)$.}

\label{fig:fig4MIMOWTCIRSimpact}
\end{figure}

\section{Numerical Analysis}

In this section, we describe the channel modeling, and present numerical
results and discussions.

\subsection{Channel Modeling}

We consider the scenario where the small-scale fading for all of the
wireless links are assumed to follow Rician distribution. Therefore,
Alice-Bob and Alice-Eve links are respectively modeled as~\cite{Stefan2021MIMOIRS},
$\mathbf{H}_{AB}=\sqrt{\tfrac{\zeta_{AB}^{-1}}{\kappa+1}}\bigl(\sqrt{\kappa}\mathbf{H}_{AB,\mathrm{LOS}}+\mathbf{H}_{AB,\textrm{NLOS}}\bigr)$
and $\mathbf{H}_{AE}=\sqrt{\frac{\zeta_{AE}^{-1}}{\kappa+1}}\bigl(\sqrt{\kappa}\mathbf{H}_{AE,\mathrm{LOS}}+\mathbf{H}_{AE,\mathrm{NLOS}}\bigr)$,
where $\mathbf{H}_{AB,\mathrm{LOS}},\mathbf{H}_{AB,\mathrm{NLOS}}\in\mathbb{C}^{N_{r}\times N_{t}}$,
and $\mathbf{H}_{AE,\mathrm{LOS}},\mathbf{H}_{AE,\mathrm{NLOS}}\in\mathbb{C}^{N_{e}\times N_{t}}$.
Here $\mathbf{H}_{AB,\mathrm{LOS}}$ accounts for the line-of-sight
(LOS) components between Alice to Bob, and the elements in $\mathbf{H}_{AB,\mathrm{LOS}}$
are defined as $e^{-j2\pi l_{r,t}/\upsilon}$ where $l_{r,t}$ is
the distance between the $t$-th antenna of Alice and the $r$-th
antenna of Bob, and $\upsilon$ denotes wavelength of the transmitted
signal. Moreover, $\kappa=1$ denotes the Rician $K$ factor, and
$\mathbf{H}_{AB,\mathrm{NLOS}}\sim\mathcal{CN}\left(\mathbf{0},\mathbf{I}\right)$
accounts for the non-line-of-sight (NLOS) components between Alice
and Bob. Similarly, $\mathbf{H}_{AE,\mathrm{LOS}}$ accounts for the
LOS components between Alice and Eve, and the elements in $\mathbf{H}_{AE,\mathrm{LOS}}$
are defined as $e^{-j2\pi l_{e,t}/\upsilon}$ where $l_{e,t}$ is
the distance between the $t$-th antenna of Alice and the $e$-th
antenna of Eve, and $\mathbf{H}_{AE,\mathrm{NLOS}}\sim\mathcal{CN}\left(\mathbf{0},\mathbf{I}\right)$
accounts for the NLOS components between Alice and Eve. In this paper,
we consider $\upsilon=15$~cm, which corresponds to the carrier frequency
of 2~GHz. We note that $l_{r,t}$ and $l_{e,t}$ are calculated according
to the system model in Fig. \ref{fig:SystemModel}. Moreover, $\zeta_{AB}$
and $\zeta_{AE}$ denote the free-space path loss (FSPL) coefficients,
defined as $\zeta_{AB}\triangleq(4\pi/\upsilon)^{2}l_{AB}^{\varepsilon}$
and $\zeta_{AE}\triangleq(4\pi/\upsilon)^{2}l_{AE}^{\varepsilon}$,
respectively~\cite{Rappaport2017FSPL}. We define $l_{AB}\triangleq\sqrt{D^{2}+(l_{t}-l_{r})^{2}}$
as the distance between Alice and Bob (i.e., the distance between
($0,0,l_{t}$) and ($D,0,l_{r}$)), $l_{AE}\triangleq D_{E}^{2}+(l_{t}-l_{e})^{2}$
as the distance between Alice and Eve (i.e., the distance between
($0,0,l_{t}$) and ($D_{E},0,l_{e}$)), and $\varepsilon=3$ is the
path loss exponent.

In a similar fashion, the channel between Alice and IRS, and that
between IRS and Bob are modeled as $\mathbf{H}_{AI}=\sqrt{\frac{1}{\kappa+1}}(\sqrt{\kappa}\mathbf{H}_{AI,\mathrm{LOS}}+\mathbf{H}_{AI,\mathrm{NLOS}})$
and $\mathbf{H}_{IB}=\sqrt{\frac{\zeta_{IB}^{-1}}{\kappa+1}}(\sqrt{\kappa}\mathbf{H}_{IB,\mathrm{LOS}}+\mathbf{H}_{IB,\mathrm{NLOS}})$,
respectively, where $\mathbf{H}_{AI,\mathrm{LOS}},\mathbf{H}_{AI,\mathrm{NLOS}}\in\mathbb{C}^{N\times N_{t}}$,
and $\mathbf{H}_{IB,\mathrm{LOS}},\mathbf{H}_{IB,\mathrm{NLOS}}\in\mathbb{C}^{N_{r}\times N}$.
The elements in $\mathbf{H}_{AI,\mathrm{LOS}}$ are defined as $e^{-j2\pi l_{i,t}/\upsilon}$
with $l_{i,t}$ being the distance between the $t$-th transmit antenna
of Alice and the $i$-th reflecting plate of IRS, and $\mathbf{H}_{AI_{\mathrm{NLOS}}}\sim\mathcal{CN}$$\left(\mathbf{0},\mathbf{I}\right)$.
Analogously, the elements in $\mathbf{H}_{IB,\mathrm{LOS}}$ are defined
as $e^{-j2\pi l_{r,i}/\upsilon}$ with $l_{r,i}$ denoting the distance
between the $r$-th receiver antenna of Bob and the $i$-th reflecting
element of the IRS and $\mathbf{H}_{IB,\mathrm{NLOS}}\sim\mathcal{CN}\left(\mathbf{0},\mathbf{I}\right)$.
The FSPL coefficient $\zeta_{IB}$ is modeled as~(c.f.~\cite{danufane2020pathloss})
$\zeta_{IB}=\frac{256\pi^{2}\upsilon^{-4}\mathcal{D}_{t}^{2}\mathcal{D}_{r}^{2}}{\left((l_{t}/\mathcal{D}_{t})+(l_{r}/\mathcal{D}_{r})\right)^{2}}$,
where $\mathcal{D}_{t}\triangleq\sqrt{(D/2)^{2}+l_{t}^{2}}$ and $\mathcal{D}_{r}\triangleq\sqrt{(D/2)^{2}+l_{r}^{2}}$.
Following a similar line of arguments, the channel between IRS and
Eve is modeled as $\mathbf{H}_{IE}=\sqrt{\frac{\zeta_{IE}^{-1}}{\kappa+1}}(\sqrt{\kappa}\mathbf{H}_{IE,\mathrm{LOS}}+\mathbf{H}_{IE,\mathrm{NLOS}}),$where
$\mathbf{H}_{IE,\mathrm{LOS}},\mathbf{H}_{IE,\mathrm{NLOS}}\in\mathbb{C}^{N_{e}\times N}$.
The elements in $\mathbf{H}_{IE,\mathrm{LOS}}$ are defined as $e^{-j2\pi l_{e,i}/\upsilon}$
with $l_{e,i}$ being the distance between the $e$-th eavesdrop antenna
and $i$-th reflecting plate of IRS, and $\mathbf{H}_{IE,\mathrm{NLOS}}\sim\mathcal{CN}\left(\mathbf{0},\mathbf{I}\right)$.
Moreove, the FSPL coefficient for IRS-Eve links is expressed as $\zeta_{IE}=\frac{256\pi^{2}\upsilon^{-4}\mathcal{D}_{t}^{2}\mathcal{D}_{e}^{2}}{\left(\left(l_{t}/\mathcal{D}_{t}\right)+\left(l_{e}/\sqrt{D_{E}+l_{e}^{2}}\right)\right)^{2}}$,
where $\mathcal{D}_{e}\triangleq\sqrt{(D_{E}/2)^{2}+l_{e}^{2}}$.

\subsection{Numerical Results}

In this subsection, we provide numerical results to evaluate the performance
of Algorithm \ref{alg:BSM}, as well as to show the effect of different
parameters of interest on the achieved secrecy rate of the IRS-assisted
MIMOME system under consideration. The numerical experiments are performed
using MATLAB (R2109a) on a 64-bit Windows machine with 16~GB RAM
and an Intel Core i7 3.20 GHz processor. Moreover, for all the figures,
we assume $\sigma_{n}^{2}=-95$ dBW, $h_{T}=3$~m, $h_{R}=2.5$~m,
$h_{E}=2$~m, $h_{I}=5$~m, $l_{t}=20$~m, $l_{r}=15$~m, $l_{e}=35$~m,
$\mathscr{\iota}_{a}=0.05$~m, $\mathscr{\iota}_{b}=0.25$~m, $\mathscr{\iota}_{e}=0.03$~m,
$l_{i}=0.03$~m, $D=50$~m and $D_{e}=40$~m. We also assumed that
all of the reflecting element at the IRS to be a square of size $0.01$~m
$\times$ $0.01$~m, and the gap between each reflecting element
to be 0.01~m.

In Fig.~\ref{fig2Convergence}, we show the convergence performance
of ~Algorithm~\ref{alg:BSM} for one set of randomly generated channels.
We also plot the convergence of~\cite[Algorithm 3]{Limeng2020WTCIRS}
for comparison. It can be noted from the figure that our BSM-based
proposed method requires a comparable number of iterations to converge,
compared to the algorithm used in~\cite{Limeng2020WTCIRS}. In particular,
\cite[Algorithm 3]{Limeng2020WTCIRS} can achieve higher secrecy rates
for some initial iterations which is explained by the fact that the
optimization of $\mathbf{X}$ is done exactly, while Algorithm~\ref{alg:BSM}
only optimizes a lower bound of the secrecy rate. However, both methods
achieves the same secrecy rate at the convergence.

The main benefit of Algorithm~ \ref{alg:BSM} is closed-form designs
in each iteration, which eventually lead to much lower run-time to
compute a solution. This point is clearly demonstrated in Fig.~\ref{figThetaTimeCheck}
where we compare the average run time of~Algorithm~\ref{alg:BSM}
and~\cite[Algorithm 3]{Limeng2020WTCIRS}. It is clearly evident
from the figure that our closed-form-based proposed algorithm to find
a stationary solution to the secrecy maximization problem under consideration
requires significantly less time compared to the existing benchmark
solution, which establishes the superiority of our proposed solution. \footnotetext {We thank the authors of \cite{Limeng2020WTCIRS} for sharing the source code for their barrier method}

In Fig.~\ref{fig3IRSimpact}, we show the effect of increasing the
number of reflecting elements, i.e., $N$, on the average secrecy
rate of the system for different number of antennas at Eve, i.e.,
$N_{e}$. The average secrecy rates are obtained for $10^{3}$ channel
realizations. The benefit of using the IRS is clearly evident from
the figure, as the system with IRS achieves a notably higher average
secrecy rate compared to the ones without any IRS. It can also be
observed from the figure that for a fixed value of $N$, the secrecy
rate of the system reduces when the number of antennas at Eve increases
since the secure degree-of-freedom for Bob decreases accordingly.
However, for a fixed value of $N_{e}$, an increase in the value of
$N$ results in a significant increase in the secrecy rate of the
system. This occurs because when the number of reflecting plates at
IRS is large, the IRS can perform highly-focused passive beamforming
towards Bob to enhance the secrecy performance of the system.

In Fig.~\ref{fig:fig4MIMOWTCIRSimpact}, we compare the average achievable
secrecy rate of the IRS-assisted MIMOME WTC system with that of the
ones without IRS, for different values of the transmit power from
Alice. According to~\cite{MIMOME_WTC}, as $P_{0}\to\infty$, the
slope of the average secrecy rate approaches zero if $\mathrm{rank}\bigl(\mathbf{H}_{E}(\boldsymbol{\theta})\bigr)=N_{t}$,
and thus the average secrecy rate is expected to saturate. The purpose
of this numerical experiment is to understand how the IRS can improve
this saturation point. We remark that $\mathrm{rank}\bigl(\mathbf{H}_{E}(\boldsymbol{\theta})\bigr)=N_{t}$
for both the cases considered in Fig.~\ref{fig:fig4MIMOWTCIRSimpact},
i.e., $N_{e}=8$ and $N_{e}=12$. It can be observed clearly from
Fig.~\ref{fig:fig4MIMOWTCIRSimpact} that beyond a certain value
of $P_{0}$, the slope of the average secrecy rate starts decreasing,
which will eventually lead to a saturation in the achieved secrecy
rate for large enough values of $P_{0}$. The important observation
is that the saturated value for an IRS-assisted system is significantly
larger that that of the system without IRS, which clearly establishes
the superiority of the IRS-assisted systems even in the large transmit
power regimes.

\section{Conclusion}

In this paper, we have proposed an efficient numerical method to maximize
the achievable secrecy rate for an IRS-assisted Gaussian MIMOME WTC
system. We have used a block successive maximization method to jointly
optimize the transmit covariance matrix and the IRS phase shifts.
The obtained results have confirmed a faster convergence and lower
complexity of the proposed method compared to an existing solution
which uses a combination of barrier method and bisection search. Furthermore,
our results have also demonstrated the superiority of IRS-assisted
systems over those without IRS, including a significantly higher achievable
secrecy rate of the former in the high transmit power regime.

\appendices{}

\balance

\section{\label{sec:APP-Closed-form-solution}Expressions for $\mathbf{P}_{i}$,
$\mathbf{Q}_{i}$ $\mathbf{R}_{i}$, and $\mathbf{S}_{i}$ in (\ref{eq:EqPQRS})}

To obtain the expressions for $\mathbf{P}_{i}$, $\mathbf{Q}_{i}$,
$\mathbf{R}_{i}$, and $\mathbf{S}_{i}$ in (\ref{eq:EqPQRS}) we
simply group the involved matrices properly. Specifically, let $\hat{\mathbf{H}}_{AB}=\bar{\mathbf{H}}_{AB}\mathbf{X}^{1/2}$,
$\hat{\mathbf{H}}_{AE}=\bar{\mathbf{H}}_{AE}\mathbf{X}^{1/2}$, and
$\hat{\mathbf{H}}_{AI}=\mathbf{H}_{AI}\mathbf{X}^{1/2}$. Then, following~\cite{Zhang2020MIMOIRS},
we can write $\mathbf{P}_{i}$, $\mathbf{Q}_{i}$, $\mathbf{R}_{i}$
and $\mathbf{S}_{i},$ respectively, as\small{\begin{subequations}
\begin{align}
\mathbf{P}_{i} & =\mathbf{I}+\Bigl(\hat{\mathbf{H}}_{AB}+\sum\nolimits _{j\in\{\mathcal{I}\setminus i\}}\theta_{j}\bar{\mathbf{h}}_{j}\hat{\mathbf{h}}_{j}\herm\Bigr)\nonumber \\
 & \!\!\!\times\Bigl(\hat{\mathbf{H}}_{AB}+\sum\nolimits _{j\in\{\mathcal{I}\setminus i\}}\theta_{j}\bar{\mathbf{h}}_{j}\hat{\mathbf{h}}_{j}\herm\Bigr)\herm+\bar{\mathbf{h}}_{i}\hat{\mathbf{h}}_{i}\herm\hat{\mathbf{h}}_{i}\bar{\mathbf{h}}_{i}\herm,\\
\mathbf{Q}_{i} & =\bar{\mathbf{h}}_{i}\hat{\mathbf{h}}_{i}\herm\Bigl(\hat{\mathbf{H}}_{AB}\herm+\sum\nolimits _{j\in\{\mathcal{I}\setminus i\}}\theta_{j}^{*}\hat{\mathbf{h}}_{j}\bar{\mathbf{h}}_{j}\herm\Bigr),\\
\mathbf{R}_{i} & =\mathbf{I}+\Bigl(\hat{\mathbf{H}}_{AE}+\sum\nolimits _{j\in\{\mathcal{I}\setminus i\}}\theta_{j}\tilde{\mathbf{h}}_{j}\hat{\mathbf{h}}_{j}\herm\Bigr)\nonumber \\
 & \times\Bigl(\hat{\mathbf{H}}_{AE}+\sum\nolimits _{j\in\{\mathcal{I}\setminus i\}}\theta_{j}\tilde{\mathbf{h}}_{j}\hat{\mathbf{h}}_{j}\herm\Bigr)\herm+\tilde{\mathbf{h}}_{i}\hat{\mathbf{h}}_{i}\herm\hat{\mathbf{h}}_{i}\tilde{\mathbf{h}}_{i}\herm,\\
\mathbf{S}_{i} & =\tilde{\mathbf{h}}_{i}\hat{\mathbf{h}}_{i}\herm\Bigl(\hat{\mathbf{H}}_{AE}\herm+\sum\nolimits _{j\in\{\mathcal{I}\setminus i\}}\theta_{j}^{*}\hat{\mathbf{h}}_{j}\tilde{\mathbf{h}}_{j}\herm\Bigr),
\end{align}
 \end{subequations}}where $\hat{\mathbf{h}}_{i}$, $\bar{\mathbf{h}}_{i}$
and $\tilde{\mathbf{h}}_{i}$ are the $i$-th column of $\hat{\mathbf{H}}_{AI}\herm$,
$\mathbf{H}_{IB}$ and $\mathbf{H}_{IE}$, respectively. Note that
we can equivalently rewrite $\mathbf{P}_{i}$ and $\mathbf{Q}_{i}$
as
\begin{subequations}
\begin{align}
\mathbf{P}_{i} & =\mathbf{I}+\bigl(\mathbf{P}-\theta_{i}\bar{\mathbf{h}}_{i}\hat{\mathbf{h}}_{i}\herm\bigr)\bigl(\mathbf{P}\herm-\theta_{i}^{\ast}\hat{\mathbf{h}}_{i}\bar{\mathbf{h}}_{i}\herm\bigr)+\bar{\mathbf{h}}_{i}\hat{\mathbf{h}}_{i}\herm\hat{\mathbf{h}}_{i}\bar{\mathbf{h}}_{i}\herm\label{eq:Pi:simple}\\
\mathbf{Q}_{i} & =\bar{\mathbf{h}}_{i}\hat{\mathbf{h}}_{i}\herm\Bigl(\mathbf{P}\herm-\theta_{i}^{\ast}\hat{\mathbf{h}}_{i}\bar{\mathbf{h}}_{i}\herm\Bigr),
\end{align}
\end{subequations}
where $\mathbf{P}=\hat{\mathbf{H}}_{AB}+\sum\nolimits _{j\in\{\mathcal{I}\}}\theta_{j}\bar{\mathbf{h}}_{j}\hat{\mathbf{h}}_{j}\herm$.

\section{\label{sec:Proof-of-Proposition}Proof of Lemma~\ref{lemma:converge}}

First, from Line \ref{alg:BSM:Xupdate} of Algorithm \ref{alg:BSM}
we have
\begin{multline}
C_{s}(\boldsymbol{\theta}_{k},\mathbf{\mathbf{X}}_{k})\geq\hat{C}_{s}(\boldsymbol{\theta}_{k},\mathbf{\mathbf{X}}_{k})\\
\geq\hat{C}_{s}(\boldsymbol{\theta}_{k},\mathbf{X}_{k-1})=C_{s}(\boldsymbol{\theta}_{k},\mathbf{X}_{k-1}).\label{eq:lemma:proof-1}
\end{multline}
Note that the first inequality is due to the fact that $\hat{C}_{s}(\boldsymbol{\theta}_{k},\mathbf{\mathbf{X}}_{k})$
is a lower bound on the secrecy rate, the second inequality is because
$\mathbf{X}_{k}=\argmax_{\mathbf{X}\in\mathcal{X}}\hat{C}_{s}(\boldsymbol{\theta}_{k},\mathbf{X})$
and the optimal objective is no less than the objective at a feasible
point, and the equality is obvious from (\ref{eq:ConvX1}). Let $\boldsymbol{\theta}_{k}\triangleq[\theta_{k,1},\theta_{k,2},\ldots,\theta_{k,N}]\trans$,
then Line~\ref{alg:BSM:thetaupdate} of Algorithm~\ref{alg:BSM}
implies the following sequence of inequalities. 
\begin{align}
C_{s}\bigl(\boldsymbol{\theta}_{k},\mathbf{\mathbf{X}}_{k-1}\bigr) & =C_{s}\bigl([\theta_{k,1},\theta_{k,2},\ldots,\theta_{k,N}]\trans,\mathbf{\mathbf{X}}_{k-1}\bigr)\nonumber \\
 & \hspace{-60pt}\geq C_{s}\bigl([\theta_{k,1},\theta_{k,2},\ldots,\theta_{k,N-1},\theta_{k-1,N}],\mathbf{\mathbf{X}}_{k-1}\bigr)\nonumber \\
 & \hspace{-60pt}\geq C_{s}\bigl([\theta_{k,1},\theta_{k,2},\ldots,\theta_{k-1,N-1},\theta_{k-1,N}]\trans,\mathbf{\mathbf{X}}_{k-1}\bigr)\nonumber \\
 & \hspace{-60pt}\geq C_{s}\bigl([\theta_{k-1,1},\theta_{k-1,2},\ldots,\theta_{k-1,N-1},\theta_{k-1,N}],\mathbf{\mathbf{X}}_{k-1}\bigr)\nonumber \\
 & \hspace{-60pt}=C_{s}\bigl(\boldsymbol{\theta}_{k-1},\mathbf{\mathbf{X}}_{k-1}\bigr).\label{eq:SecondHalf}
\end{align}
Therefore, combining~(\ref{eq:lemma:proof-1}) and~(\ref{eq:SecondHalf}),
we achieve~(\ref{eq:CSequence}).

The second part of Lemma~\ref{lemma:converge} can be proved as follows.
Note the the objective is bounded from above due to the power constraint.
Due to~(\ref{eq:CSequence}), the objective sequence is convergent.
i.e. $\lim_{k\to\infty}C_{s}(\boldsymbol{\theta}_{k},\mathbf{\mathbf{X}}_{k})=C_{s}(\boldsymbol{\theta}_{\ast},\mathbf{\mathbf{X}}_{\ast})$.
Since the feasible set is compact, there exists a subsequence $(\boldsymbol{\theta}_{k_{j}},\mathbf{\mathbf{X}}_{k_{j}})$
converging to $(\boldsymbol{\theta}_{\ast},\mathbf{\mathbf{X}}_{\ast})$.
The proof that $(\boldsymbol{\theta}_{\ast},\mathbf{\mathbf{X}}_{\ast})$
is a stationary point of~(\ref{eq:MainProblem}) follows the arguments
in~\cite{razaviyayn2012BSM}, which are skipped here the the sake
of brevity.

\bibliographystyle{IEEEtran}
\bibliography{IEEEabrv,paper}

\end{document}